\documentclass[aps,prl,twocolumn,groupedaddress]{revtex4-2}
\usepackage{units}
\usepackage{amsmath}
\usepackage{amssymb,amsfonts}
\usepackage{graphicx}
\usepackage{bm}
\usepackage{multirow,color,relsize,microtype}%ulem
\usepackage{dcolumn}% Align table columns on decimal point
\usepackage{xcolor}

\newcommand{\be}{\begin{equation}}
\newcommand{\ee}{\end{equation}}

\newcommand{\re}[1]{\text{Re}\left[#1\right]}

\begin{document}

\title{Non-Hermitian systems with a real spectrum and selective skin effect}

\author{Li Ge}
\email{li.ge@csi.cuny.edu}
\affiliation{\textls[-18]{Department of Physics and Astronomy, College of Staten Island, CUNY, Staten Island, NY 10314, USA}}
\affiliation{The Graduate Center, CUNY, New York, NY 10016, USA}

\begin{abstract}
In this work we first show a simple approach to constructing non-Hermitian Hamiltonians with a real spectrum, which are \textit{not} obtained by a non-unitary transformation such as the imaginary gauge transformation. They are given, instead, by the product of a Hermitian Hamiltonian $H_0$ and a positive semi-definite matrix $A$. Depending on whether $A$ has zero eigenvalue(s), the resulting $H$ can possess an exceptional point at zero energy. When $A$ is only required to be Hermitian instead, the resulting $H$ is  pseudo-Hermitian that can have real and complex conjugate energy levels. 
In the special case where $A$ is diagonal, we compare our approach to an imaginary gauge transformation, which reveals a selective non-Hermitian skin effect in our approach, i.e., only the zero mode is a skin mode and the non-zero modes reside in the bulk. We further show that this selective non-Hermitian skin mode has a much lower lasing threshold than its counterpart in the standard non-Hermitian skin effect with the same spatial profile, when we pump at the boundary where they are localized. 
The form of our construction can also be found, for example, in dynamical matrices describing coupled frictionless harmonic oscillators with different masses. 
\end{abstract}

\date{\today}

%\keywords{}

\maketitle

\section{Introduction}

Despite the Hermiticity of quantum mechanics, many quantum and wave systems can be described using a non-Hermitian framework, where a vast subsystem is considered as the reservoir or environment that exchanges particles and energy with the rest of the system. 
Early investigations of non-Hermitian systems in nuclear physics \cite{Gamow} employed an open boundary condition of the Schr\"odinger equation, which leads to an effectively non-Hermitian Hamiltonian and complex energy levels, with the latter known as resonances or quasi-bound states. Such states correspond to the poles of the scattering matrix, where one or more eigenvalues of the scattering matrix diverge. As such, their presence can be easily detected in the transmission and reflection spectra, especially when the resonances are close to the real axis. This distance gives the decay rate of a resonance, and it can be compensated by injecting energy into the system, as demonstrated in masers and lasers. For example, one can move a resonance in an optical system to the real frequency axis by providing optical gain, leading to a lasing mode \cite{SALT_Science,SALT_PRA}.

The combination of these two elements of non-Hermiticity, i.e., loss and gain, can even lead to an entirely real spectrum of a non-Hermitian Hamiltonian, if the complex potential satisfies parity-time (PT) symmetry \cite{Bender}. This finding hence raised the possibility of a non-Hermitian extension of standard quantum mechanics. Such PT-symmetric systems can be shown to have pseudo-Hermiticity \cite{Mostafazadeh2}, meaning that the Hamiltonian and its Hermitian conjugate are related by a similar transformation. However, just being PT-symmetric or pseudo-Hermitian does not guarantee a real spectrum. In fact, the notion of a spontaneous PT symmetry breaking, signaled by the transition from real eigenvalues to complex conjugate ones, has inspired a cornucopia of findings in photonics and related fields \cite{NPReview,RMP,Longhi}, including the enhanced spectral response at an exceptional point (EP) \cite{EPsensing1,EPsensing2,EPsensing3}. Unlike a degeneracy in Hermitian systems, an EP is marked by the coalescence of both the energy levels and their wavefunctions. The energy dispersion at an EP usually diverges in some direction(s) of the parameter space, although a linear dispersion can also be found, e.g., at a Dirac EP \cite{DiracEP_PRL,DiracEP_PRB}. 

Another approach to constructing a non-Hermitian Hamiltonian with a real spectrum is to perform a non-unitary transformation of a Hermitian Hamiltonian \cite{Martinez}. Because a similar transformation leaves the spectrum unchanged, this approach guarantees an entirely real energy spectrum, and one can show that this non-Hermitian Hamiltonian is also pseudo-Hermitian. One particular family of transformations within this approach that has generated fast-growing interest is known as an imaginary gauge transformation \cite{Hatano,Hatano2}. This transformation can be understood as replacing the phase in the standard gauge transformation of Maxwell's equations with its imaginary counterpart, leading to non-reciprocal couplings as well as the superposition of an exponential envelope on all eigenstates of the system. 
As such, this transformation induces the so-called non-Hermitian skin effect: most (if not all) modes are localized at one side or corner of the system, which has been observed in photonic \cite{Gao}, acoustic \cite{acoustic}, and mechanical \cite{mechanical} systems.
This effect is not limited to systems in position space with nearest-neighbor couplings and can be extended to synthetic dimensions with longer-range couplings \cite{DiracEP_PRL,DiracEP_PRB}.

More generally, a theorem states \cite{Mostafazadeh2} that a non-Hermitian Hamiltonian $H$ must be pseudo-Hermitian in order to have a real spectrum, which is proved by assuming a biorthogonal set of eigenstates. This theorem itself, however, does not offer an explicit approach to constructing such a non-Hermitian Hamiltonian, particularly if we want to go beyond the direct non-unitary transformation mentioned above. 

In this article, we first introduce a simple approach that gives non-Hermitian Hamiltonians in their matrix forms with a real spectrum. We start with an arbitrary Hermitian Hamiltonian $H_0$, and we show that its product with a positive semi-definite matrix $A$ leads to a non-Hermitian Hamiltonian $H$ with a real spectrum. Because $A$ is arbitrary itself and this procedure is not a similar transformation, the same $H_0$ can lead to an infinite set of $H$'s with different real spectra. Depending on whether $A$ has zero eigenvalue(s), the resulting $H$ can possess an EP at zero energy. When $A$ is only required to be Hermitian instead, the resulting $H$ is pseudo-Hermitian that can have real and complex conjugate energy levels. 

In the special case where $A$ is diagonal, we compare our approach to an imaginary gauge transformation, and we show that our approach leads to a selective non-Hermitian skin effect, where only the zero mode is a skin mode and the non-zero modes reside in the bulk. Furthermore, we find that this selective non-Hermitian skin mode has a much lower lasing threshold than its counterpart in the standard non-Hermitian skin effect with the same spatial profile, when we pump at the boundary where they are localized. 
Finally, we show that the form of our construction can also be found, for example, in non-Hermitian dynamical matrices describing coupled frictionless harmonic oscillators with different masses.  

\section{Results}

\subsection{Real spectrum}

Below we start our discussion by requiring $A$ to be Hermitian but not necessarily positive semi-definite, and we show that the resulting matrix $H=H_0A$ is pseudo-Hermitian that can have either real or pairs of complex conjugate eigenvalues. For this purpose, we first note that $H^\dagger$ is simply given by $AH_0$ under this condition. Next, if $\psi_\mu$ is a right eigenstate of $H$, we then find
\begin{equation}
H^\dagger (A\psi_\mu) = AH_0(A\psi_\mu) =  AH\psi_\mu = \omega_\mu(A\psi_\mu).\label{eq:1}
\end{equation}
If $A\psi_\mu=0$, $\psi_\mu$ is then an eigenstate of $A$ (and $H$) with $\omega_\mu=0$ (which is real). If $A\psi_\mu\neq0$, there is then one left eigenstate $\tilde{\psi}_\nu^T$ of $H$, defined by $\tilde{\psi}_\nu^T H = \omega_\nu\tilde{\psi}_\nu^T$, or equivalently,
\be
H^\dagger \tilde{\psi}_\nu^*  = [\tilde{\psi}_\nu^T H]^\dagger = \omega_\nu^* \tilde{\psi}_\nu^*,
\ee
that satisfies
\be
\omega_\mu = \omega_\nu^*,\quad A\psi_\mu = \tilde{\psi}_\nu^*. \label{eq:pseudoH}
\ee
$\mu,\nu$ are not necessarily the same, and only when they are does $H$ have a real spectrum. These spectral features indicate that $H$ is pseudo-Hermitian, and if $H_0$ is invertible, the ``metric'' $\eta$ in the definition of pseudo-Hermiticity, i.e., $H^\dagger = \eta H \eta^{-1}$, is simply given by $H_0^{-1}$:
\be
H_0^{-1} H H_0 = H_0^{-1} (H_0A) H_0 = AH_0 = H^\dagger.
\ee
%We do not give $\eta$ explicitly when $H_0$ is not invertible but refer to the first relation in Eq.~(\ref{eq:pseudoH}) as the working definition of pseudo-Hermiticity in this case.

Now, if $A$ is not just Hermitian but also positive-definite, we can write $A$ as $B^\dagger B$ where $B$ is an arbitrary square matrix of full rank. We then observe that the biorthogonal inner product 
\be
\tilde{\psi}_\nu^T\psi_\mu = (\tilde{\psi}_\nu^*)^\dagger\psi_\mu = (A\psi_\mu)^\dagger\psi_\mu = (B\psi_\mu)^\dagger (B\psi_\mu) \geq 0,\label{eq:innerProduct}
\ee
where we have used the second relation in Eq.~(\ref{eq:pseudoH}). We note that the equality holds only when $B\psi_\mu=0$, which contradicts that $B$ is of full rank. Therefore, $\tilde{\psi}_\nu^T\psi_\mu$ must be finite.

Away from an EP, the biorthogonal relation $\tilde{\psi}_{\mu'}^T\psi_\mu=\delta_{\mu'\mu}$ holds. At an EP, its left and right wave functions $\tilde{\psi}^T_\mu, \psi_\mu$ are also ``self-orthogonal,'' i.e., they satisfy $\tilde{\psi}_{\mu}^T\psi_\mu=0$, while $\tilde{\psi}_{\mu'}^T\psi_\mu=0$ still holds when $\mu'$ is different from $\mu$. Therefore, if an inner product $\tilde{\psi}_{\nu}^T\psi_\mu$ is finite, then two conditions must be satisfied: (1) $\psi_\mu$ is an eigenstate of $H$ away from an EP, and (2) $\mu,\nu$ must be the same index. Since the inner product $\tilde{\psi}_{\nu}^T\psi_\mu$ is finite for any right eigenstate $\psi_\mu$ of $H$, then none of the right eigenstates of $H$ is at an EP, which indicates that $H$ is not defective and has no EPs. 

More importantly, with a positive-definite $A$ and $\mu=\nu$, the first relation in Eq.~(\ref{eq:pseudoH}) then tells us that 
\be
\omega_\mu = \omega_\mu^* \in\mathbb{R}\label{eq:real}
\ee
for all $\psi_\mu$'s, which concludes our proof that all energy levels of $H$ are real and away from an EP when $A$ is positive-definite.

If $A$ is allowed to be positive semi-definite and have zero eigenvalue(s), we can still write $A=B^\dagger B$ but $B$ now has zero eigenvalue(s). As a result, Eq.~(\ref{eq:innerProduct}) still holds but the equality cannot be eliminated, i.e., with $B\psi_\mu=0$, and in turn, $A\psi_\mu=0$, which violates the condition under which Eq.~(\ref{eq:innerProduct}) is derived. As a result, we find the following properties: 

\begin{enumerate}
\item If $\psi_\mu$ is an eigenstate of $H$ with a non-zero eigenvalue $\omega_\mu$, then $B\psi_\mu\neq0$, and in turn, we find that $\psi_\mu$ is not at an EP and $\omega_\mu\in\mathbb{R}$, similar to the case where $A$ is positive-definite.
\item Even if $\psi_\mu$ is an eigenstate of $H$ with a zero eigenvalue, it is still not at an EP as long as $B\psi_\mu\neq0$.
\item $\psi_\mu$ can be at an EP only if it is an eigenstate of $B$ (and $A$, $H$) with a zero eigenvalue. In other words, $H$ can have an EP only at $\omega_\mu=0$, with the necessary condition that the corresponding eigenstate $\psi_\mu$ satisfies $A\psi_\mu=0$.
\end{enumerate}

%we still find $\omega_\mu\in\mathbb{R}$ for all $\psi_\mu$'s that are not eigenstates of $A$ with a zero eigenvalue, but now there is no guarantee that a zero eigenstate of $A$ (and $H$) has a finite inner product with the corresponding left eigenstate of $H$. Consequently, $\omega_\mu=0$ has the possibility of being an EP of $H$. 

In all our discussions above, the order of $A$ and $H_0$ in our construction of $H$ is inconsequential; the other order $AH_0$ gives $H^\dagger$ with the same real eigenvalues.

To understand the conditions leading to the real spectrum of $H$ in our construction, we focus on the special case where $A$ is diagonal and $H_0$ is given by a tight-binding model:
\be
H_0 = \sum_{j} \omega_j|j\rangle\langle j| \,+\, t_{j+1,j}|j+1\rangle\langle j| + t_{j,j+1}|j\rangle\langle j+1|, \label{eq:TBM}
\ee
where $\omega_j$ is the on-site potential and $t_{j+1,j}$ is the coupling from lattice site $j$ to $j+1$ and equals the coupling in the opposite direction, i.e., $t_{j,j+1}$. By denoting the diagonal elements of $A$ by $a_j$'s, we first change the potential landscape, from $\omega_j$'s in $H_0$ to $a_j\omega_j$ in $H$, which does not take place in an imaginary gauge transformation \cite{chiral}. Although this procedure maintains the sign of the local potential, it can change a potential barrier to a well and vice versa. At the same time, we also scale the two couplings $t_{j\pm1,j}$ from the site $j$ by the same $a_j$. This scaling, though, is direction dependent: the pair of couplings $t_{j,j+1},t_{j+1,j}$ between the two sites $j,j+1$ are scaled by $a_{j+1},a_j$, respectively. 

\begin{figure}[b]
\includegraphics[clip,width=\linewidth]{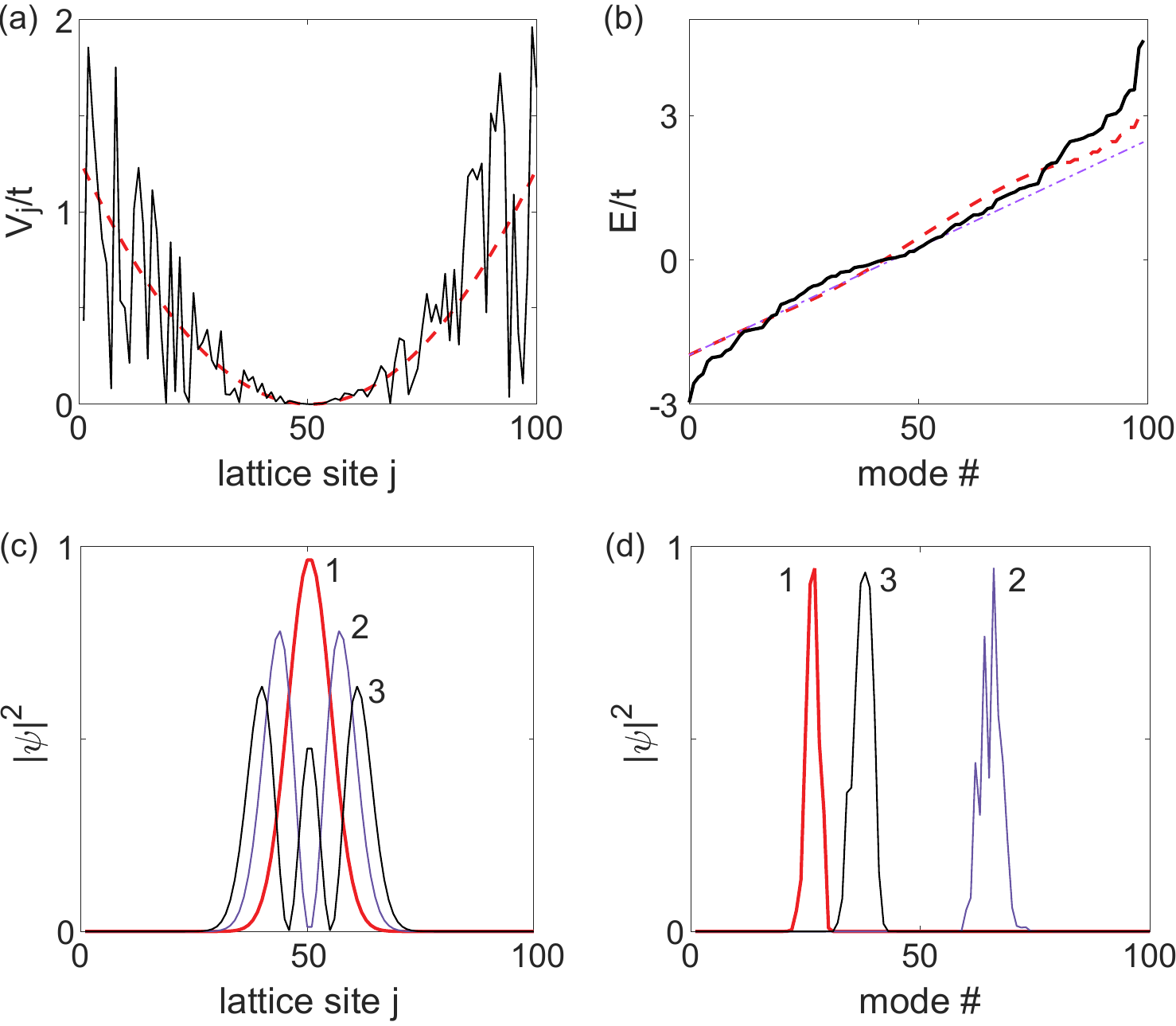}
\caption{\textbf{An example of constructing a non-Hermitian system with a real spectrum.} (a) Potential in $H$ (solid) and $H_0$ (dashed). $\omega^2=t/1000.$ (b) Spectra of $H$ and $H_0$, with that of the latter compared with the approximation given by Eq.~(\ref{eq:harmonic}) (dash-dotted). (c,d) Wave functions of the lowest three energy states in $H_0$ and $H$, respectively. } \label{fig:harmonic}
\end{figure}

As an example, below we consider a one-dimensional system $H_0$ with a harmonic potential \cite{harmonic}
\be
\omega_j=[j-(N-1)/2]^2\omega^2/2\quad(j=1,\ldots,N=100) 
\ee
[see Fig.~\ref{fig:harmonic}(a)]. The coupling $t\in\mathbb{R}$ is taken to be uniform. For the $N$ eigenstates of $H_0$, their lowest energies and corresponding wave functions follow closely the continuous case [see Figs.~\ref{fig:harmonic}(b,c)], i.e., 
\be
E_q + 2|t| \approx (q-1/2)\,\tilde{\omega},\label{eq:harmonic}
\ee
where $q$ is a small positive integer. The offset $2|t|$ is due to the discrete nature of the system, and the effective natural frequency of the harmonic potential is given by $\tilde{\omega}\equiv\omega\sqrt{2|t|}$. 

Once we scale $\omega_j$ (and $t$) by an arbitrary $a_j\in(0,2]$ [see Fig.~\ref{fig:harmonic}(a)], the spectrum of $H=H_0A$ is very different and has a much longer tail at both the high and low end of the spectrum [see Fig.~\ref{fig:harmonic}(b)]. Despite the asymmetric couplings in $H$, the wave functions of its eigenstates are still Anderson localized due to the random scaling we have introduced to the potential, especially the states near the two spectral ends where the spectral density is low [see Fig.~\ref{fig:harmonic}(d)].

Having exemplified our construction of a non-Hermitian Hamiltonian with a real spectrum, we note that the Hermitian Hamiltonian $H_0$ in our construction of $H$ cannot be replaced by a non-Hermitian matrix $H'$ with a real spectrum; the resulting $H''=H'A$ is not even pseudo-Hermitian in general. 

One obvious exception is $H'=A^{-1}H_0$ where $H_0$ is Hermitian, leading to a non-Hermitian $H''=A^{-1}H_0A$ with a real spectrum if $A$ is non-unitary. This similar transformation from $H_0$ to $H$, with $A$ being diagonal and positive-definite, is the imaginary gauge transformation mentioned previously, which introduces a position-dependent scaling to the eigenstates of $H_0$ and leads to the non-Hermitian skin effect \cite{Hatano2,Gao}.       

\subsection{Selective non-Hermitian skin effect}

\begin{figure}[b]
\includegraphics[clip,width=\linewidth]{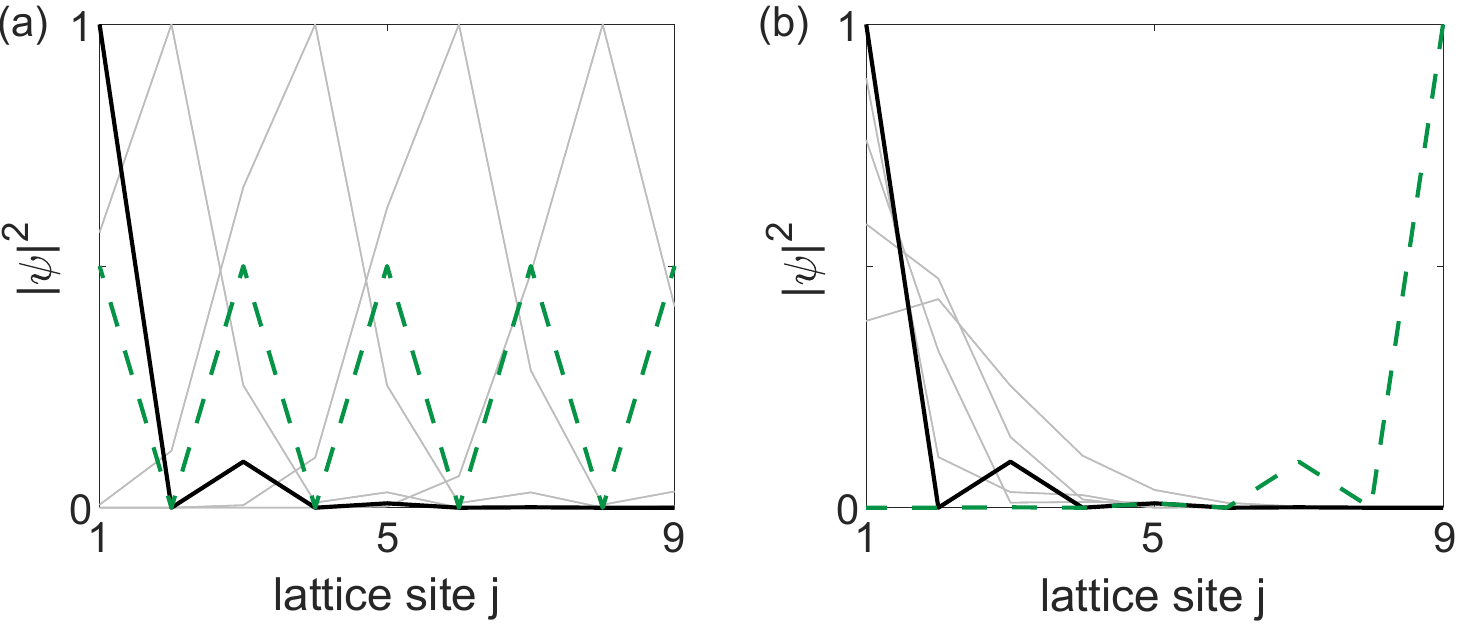}
\caption{\textbf{Selective non-Hermitian skin effect (a) versus standard non-Hermitian skin effect (b)}. Thick and thin solid lines show the zero and non-zero modes, respectively. Each thin line represents two modes related by chiral symmetry that have the same spatial profile. Dashed line in each panel shows the left eigenstate of its zero mode.}\label{fig:skin}
\end{figure}

The wave functions of $H$ and $H_0$ in our construction, on the other hand, are not simply related in general, and so are their (real) eigenvalues. This property, as it turns out, leads to a \textit{selective} non-Hermitian skin effect: only the zero mode of the system, i.e., with $\omega_\mu = 0$, is  exponentially localized at one edge of the system, while the non-zero modes remain in the bulk [see Fig.~\ref{fig:skin}(a)]. Here the original Hermitian $H_0$ is given by the tight-binding model (\ref{eq:TBM}) with $\omega_j=0$ and $t_{j,j+1}=t_{j+1,j}=t$, and it has chiral symmetry that warrants the zero mode in both the Hermitian case ($H_0$) and the non-Hermitian cases ($H$ and $H''$) \cite{chiral}. Furthermore, one can show explicitly that the zero modes of $H=H_0A$ and $H''=A^{-1}H_0A$ are exactly the same [cf. Figs.~\ref{fig:skin}(a,b)], but they are formed by different mechanisms as we show below. 

The standard non-Hermitian skin effect, as mentioned before, is caused by exponentiating the underlying Hermitian modes via a non-unitary transformation:
\be
a_j = s^{j-1},\quad\psi''_{\mu,j} = \psi^{(0)}_{\mu,j}s^{-(j-1)},\label{eq:skin}
\ee
where $j=1,2,\ldots,N$ and $\psi''_{\mu,j},\psi^{(0)}_{\mu,j}$ are the $\mu$th wave function of $H'',H_0$ at site $j$, respectively. The skin modes are localized on the left (right) of the system when $s>1$ ($0<s<1$). The ratio of each pair of (nearest-neighbor) couplings is given by $s^2$, and their geometric average remains unchanged as in $H_0$ and uniform in space. 

For the selective non-Hermitian skin mode in $H$, its formation is the combination of two effects. One is the same as the standard non-Hermitian skin effect but with the localization length doubled: the ratio of each pair of (nearest-neighbor) couplings is now given by $s$ instead of $s^2$, as can be easily checked. In the meanwhile, the geometric average of each pair (i.e., the effective coupling strength) also forms a geometric series, increasing by a factor of $s$ from $\sqrt{t_{j-1,j}t_{j,j-1}}$ to $\sqrt{t_{j,j+1}t_{j+1,j}}$. This increment of the effective coupling strength, similar to the alternate strong and weak couplings in the Su-Schrieffer-Heeger (SSH) array \cite{SSH} and the breathing Kagome lattice \cite{Kagome1}, contributes the same exponential factor to the wave function as the first mechanism. It then reduces the would-be localization length of the zero mode by half, and we recover the same zero-mode wave function given by Eq.~(\ref{eq:skin}): 
\be
\psi_{0,j} = \psi^{(0)}_{0,j}s^{-(j-1)}\label{eq:sskin}.
\ee
We also note that for the corresponding left eigenstate, each pair of couplings $t_{j,j+1}$,$t_{j+1,j}$ are effectively exchanged, and hence the first mechanism alone would localize the left eigenstate on the right edge of the system instead, similar to what happens in the standard non-Hermitian skin effect [see the dashed line in Fig.~\ref{fig:skin}(b)]. The second mechanism mentioned above is not affected by this exchange of $t_{j,j+1}$,$t_{j+1,j}$, because it is determined only by the geometric average of these two couplings. As a result, acting alone it would still localize the left eigenstate of the zero mode on the left edge of the system, and these two mechanisms hence cancel each other, causing the left eigenstate of the zero mode in the selective skin effect to be extended [see the dashed line in Fig.~\ref{fig:skin}(a)], which has the same spatial profile as the zero mode of the underlying Hermitian Hamiltonian $H_0$.

Also due to the increment of the effective coupling, the non-zero modes are now further away spectrally compared to those in $H_0$ (and $H''$). In the case shown in Fig.~\ref{fig:skin}(a), the next two eigenstates of $H$ are at $\omega_\mu=\pm2.38t$, while those in $H''$ shown in Fig.~\ref{fig:skin}(b) are at $\omega_\mu=\pm0.618t$. Due to such energy differences, the wave functions of non-zero modes in $H$ and $H_0$ cannot be related directly, and the analysis leading to Eq.~(\ref{eq:sskin}) does not apply to them, indicating the absence of non-Hermitian skin effect in these modes. We also note that the onsite potentials $\omega_j$'s, which are scaled in $H$ from $H_0$ in general but not in $H''$ as mentioned previously, do not play a role here because they are all zero. 

Another difference between the selective and standard non-Hermitian skin effects is their vastly different lasing thresholds with a localized pump. In this consideration, each site in the tight-binding model represents a laser cavity, and it is typically a micro-ring or micro-disk laser by itself on an integrated photonic platform \cite{microcavity1,microcavity2,microcavity3}. When the zero mode is the target lasing mode, intuitively one can pump the leftmost cavity in both systems, achieving the lowest lasing threshold by maximizing the spatial overlap between the pump and the zero mode \cite{anomalousTH}. 

In the tight-binding model, the radiation and material loss in each cavity are taken into account by making $\omega_j$ complex, i.e., adding a negative imaginary part $\kappa_0$. The pump compensates for this loss term by adding a positive imaginary part $\gamma_j$, and with spatially uniform pumping, all modes reach their non-interacting lasing threshold simultaneously, given by $\gamma_j=\kappa_0$. Here the non-interacting thresholds ignore the competition for gain among the possible lasing modes, and the lowest one of all modes is the actual lasing threshold with modal interaction included in this framework, which is a simplified version of the semiclassical laser theory \cite{SALT_Science,SALT_PRA}.

With the selective pumping scenario described above, however, only $\gamma_1$ in the leftmost cavity is nonzero, and the threshold of the zero mode is the lowest in both $H$ and $H''$. Although the zero mode has the same spatial profile in these two cases, their thresholds still differ significantly. Denoting this zero-mode lasing threshold by $D$ and $D''$ in these two cases, they are given by $D=1.44\kappa_0$ and $D''=4.99\kappa_0$ when the couplings are much stronger than the cavity loss ($\kappa_0=0.02t$). 

This difference is caused by the energy exchange between the system and the photonic environment at the coupling junctions \cite{anomalousTH}, which can be quantified by analyzing the dynamical equation for the intensity in each cavity:
\be
\frac{d|\psi_j|^2}{dt} = 2(\gamma_j-\kappa_0)|\psi_j|^2 + {\cal P}_{j,j+1} + {\cal P}_{j,j-1},\label{eq:dynamics}
\ee
where 
\begin{align}
{\cal P}_{j,j+1} &= it_{j,j+1}^*\psi^*_{j+1} \psi_{j}\, +\, c.c., \\
{\cal P}_{j,j-1} &= it_{j,j-1}^*\psi^*_{j-1} \psi_{j}\, +\, c.c.
\end{align}
are the inter-cavity power flows from cavity $j+1$ to $j$ and from cavity $j-1$ to $j$ \cite{Ge_PRA_2017b} respectively, and $c.c.$ stands for the complex conjugation of the first term. This equation indicates that  
\be
{\cal G}_{j,j+1} \equiv {\cal P}_{j,j+1} + {\cal P}_{j+1,j} =  i(t_{j,j+1}^*-t_{j+1,j})\psi^*_{j+1} \psi_{j}\, +\, c.c.\nonumber
\ee
is the power loss (if negative) or gain (if positive) at the coupling junction between cavities $j$ and $j+1$, which is non-zero only with asymmetric couplings (i.e., $t_{j,j+1}^*\neq t_{j+1,j}$). 

\begin{figure}[t]
\includegraphics[clip,width=\linewidth]{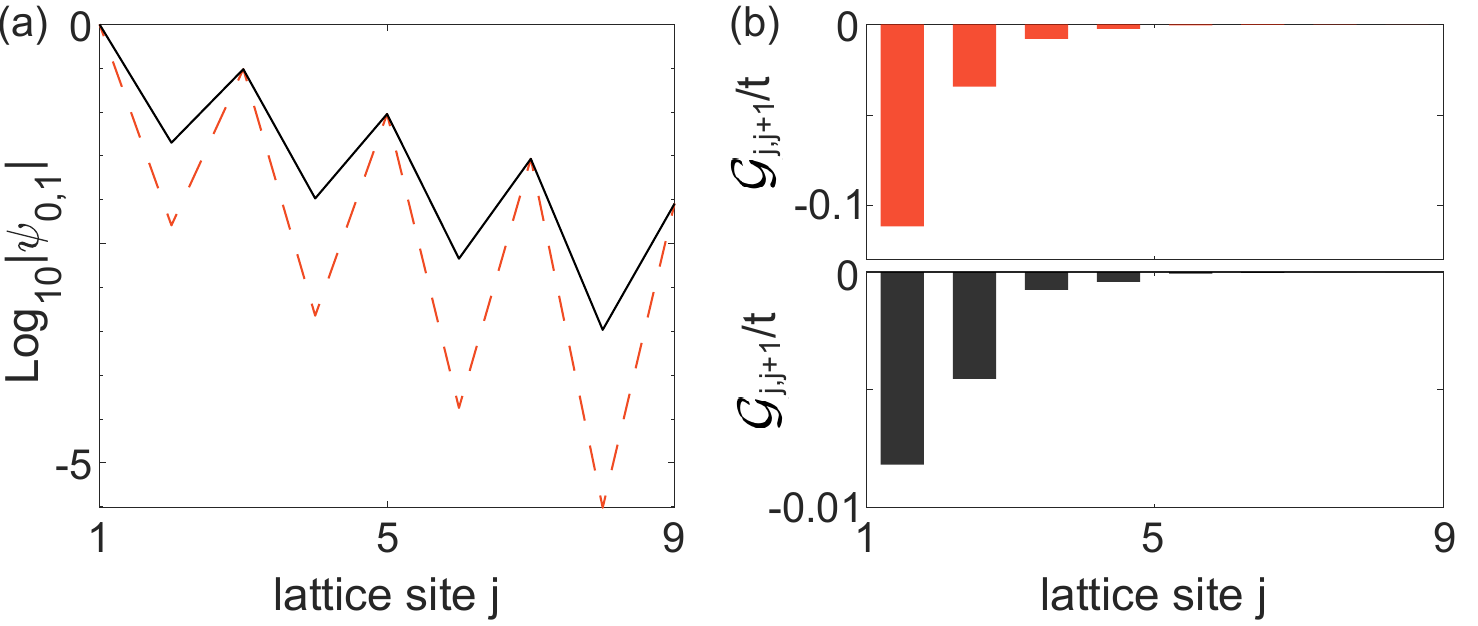}
\caption{\textbf{Analysis of the lasing mode at threshold.} (a) Spatial profile of the zero mode at the lasing threshold in $H$ (solid) and $H''$ (dashed). (b) Its loss to the photonic environment at coupling junctions in $H$ (bottom) and $H''$ (top). ${\cal G}_{j,j+1}$ is centered at $j+1/2$ and with the normalization $\psi_{0,1}=1$ in the leftmost cavity. Here $\kappa_0=0.02t$.} \label{fig:laser}
\end{figure}

Due to the non-Hermitian particle-hole (NHPH) symmetry of both $H$ and $H''$ \cite{zeromodeLaser}, the passive zero mode with $\gamma_1=0$ becomes a non-Hermitian zero mode \cite{NHFlatband_PRL,NHFlatband_PRJ} as $\gamma_1$ increases, i.e., without changing its frequency. Its spatial profile, however, is slightly perturbed from that shown in Fig.~\ref{fig:skin}, which is no longer ``dark'' with $\psi_{0,j}=0$ at the even-numbered lattice sites [Fig.~\ref{fig:laser}(a)]. This behavior can be explained using a non-Hermitian perturbation theory (see Methods) \cite{linearLoc}, and it is another requirement for ${\cal G}_{j,j+1}$'s to be non-zero in the zero mode of both $H$ and $H''$ [Fig.~\ref{fig:laser}(b)]. 
%This is another requirement for ${\cal G}_{j,j+1}$'s to be non-zero in the zero mode of both $H$ and $H''$, and their values are shown in Fig.~\ref{fig:laser}(b). 
Clearly, ${\cal G}_{j,j+1}$'s are all negative in both cases, indicating power loss at the coupling junctions. Furthermore, this power loss is one order of magnitude lower in $H$ than $H''$, leading to the much lower threshold in the former. This contrast becomes lower when the couplings are comparable to the cavity loss, e.g., we find $D = 1.35\kappa_0$ and $D''=1.62\kappa_0$ when $\kappa_0=t$.    

\begin{figure}[b]
\includegraphics[clip,width=\linewidth]{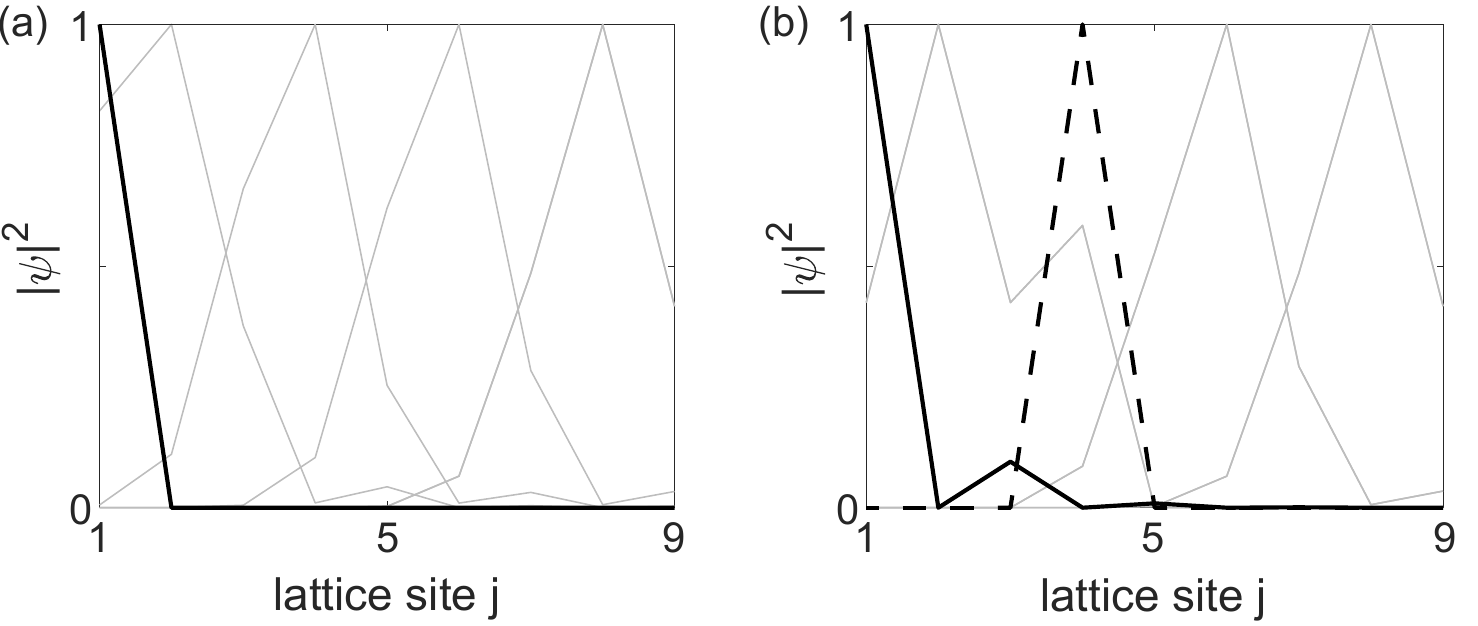}
\caption{\textbf{Zero modes and EP}. Both panels are the same as Fig.~\ref{fig:skin}(a) except for $a_1=0$ in (a) and $a_4=0$ in (b). Dashed line in (b) shows the zero mode at an EP2, while thick solid lines in (a) and (b) show their zero mode not at an EP.} \label{fig:EP}
\end{figure}

\section{Dicussions}

So far, we have considered examples with a positive-definite $A$. If $A$ has zero eigenvalue(s) (and is hence positive semi-definite), now $H$, still possessing a real spectrum, can have an EP at $\omega_\mu=0$ as mentioned previously, if the corresponding eigenstate satisfies $A\psi_\mu=0$.

As an example, we replace $a_4$ by 0 in the diagonal $A$ used in the case shown in Fig.~\ref{fig:skin}(a). The resulting $H$ has 6 real and non-zero real energy levels, paired by $\omega_\mu=-\omega_\nu$ due to the chiral symmetry and NHPH symmetry. The other three eigenvalues are zero, but they do not form an EP of order 3. One of them is accompanied by the same selective skin mode $\psi_0$ given by Eq.~(\ref{eq:sskin}), which is not an eigenstate of the modified $A$. Therefore, this eigenstate $\psi_0$ cannot be at an EP of $H$ as we explained earlier, and it forms a single-element Jordan block \cite{Jordan}. The other two zero eigenvalues form an EP of order 2 (EP2), with the coalesced eigenstate localized at $j = 4$ [Fig.~\ref{fig:EP}(b)]. This eigenstate, given by $\psi_{0'} = [0,0,0,1,0,0,0,0,0]^T$, indeed satisfies $A\psi_{0'}=0$ as required. The location of this zero mode can be easily understood, because setting $a_4=0$ means that the site $j=4$ does not couple to any other site, and the isolated wave function $\psi_{0'}$ is guaranteed to be a zero mode. The corresponding Jordan chain that completes the Hilbert space can also be found analytically, i.e., $HJ=\psi_{0'}$ with $J=[-1,0,s^{-2},0,0,0,0,0,0]^T$. The same outcome holds if we replace another $a_j$ by 0 instead, if $j$ is even.

If this index $j$ is odd instead, the outcome becomes very different. For example, if we replace $a_1$ by 0 instead, then $\omega_\mu=0$ is non-degenerate and not an EP, and the isolated wave function is localized at the left boundary [Fig.~\ref{fig:EP}(a)], replacing the selective skin mode. In this case, the zero-mode wave function $\psi_{0'}=[1,0,0,0,0,0,0,0,0]^T$ again satisfies $A\psi_{0'}=0$, but as we mentioned previously, this latter condition is a necessary but not sufficient condition for $\omega_\mu=0$ to be an EP of $H$. Furthermore, if $N$ is even instead that deprives $H_0$ of a zero mode, now $a_1=0$ gives rise to an EP of order 2, with the same isolated wave function at the left boundary (not shown). 

%The reduced density of states near the zero mode, together with being the exclusive non-Hermitian skin mode, makes the observation of the localized non-Hermitian zero mode more viable in a transmission setup.

% \vdots
% Next, instead of a random scaling matrix $A$, we turn the harmonic potential upside down. This change requires offsetting the minimum of the original $\omega_j$ by some positive value, chosen here as $2t$, i.e., $\omega_j=2t+[j-(N-1)/2]^2\omega^2/2$. The inverted potential is chosen to be $V'_j=2t-[j-(N-1)/2]^2\omega^2$, which requires 
% \be
% a_j = \frac{2t-[j-(N-1)/2]^2\omega^2}{2t+[j-(N-1)/2]^2\omega^2}
% \ee

Another way to understand these results is the following. With a positive-definite or positive semi-definite $A$, we write it again as $B^\dagger B$ where $B$ is an arbitrary square matrix. We then find
\be
(BH_0B^\dagger)(B\psi_\mu) = \omega_\mu(B\psi_\mu),
\ee
which means: 
\begin{itemize}
\item If $A$ is positive-definite, then $B$ is of full rank. Therefore, $B$ is invertible and $H$ can be obtained by a similar transformation from the Hermitian Hamiltonian $H_e\equiv BH_0B^\dagger$, i.e., 
\be
H = B^{-1}H_eB,
\ee
indicating that $H$ has an entirely real spectrum. Furthermore, $B$ provides an invertible mapping between the Hilbert spaces of $H_e$ and $H$, which means every eigenstate $\phi_\mu$ of $H_e$ is mapped to a distinct eigenstate $\psi_\mu = B^{-1}\phi_\mu$ of $H$, ruling out the possibility of $H$ having an EP. 

\item When $A$ (and $B$) is singular with zero eigenvalue(s), every eigenstate $\psi_\mu$ of $H$ is mapped to an eigenstate $\phi_\mu$ of $H_e$ with the same eigenvalue $\omega_\mu$, as long as $B\psi_\mu\neq0$. For a non-zero $\omega_\mu$, $B\psi_\mu$ cannot be zero (it leads to $\omega_\mu=0$), and hence these non-zero $\omega_\mu$'s must be real, and so is the entire spectrum of $H$. At the same time, $H_e\phi_\mu=\lambda_\mu\phi_\mu$ leads to
\be
\hspace{3mm} (B^\dagger BH_0)(B^\dagger\phi_\mu)=H^\dagger(B^\dagger\phi_\mu) = \lambda_\mu(B^\dagger\phi_\mu),
\ee
meaning every eigenstate $\phi_\mu$ of $H_e$ is mapped to an eigenstate of $H^\dagger$ with the same eigenvalue $\lambda_\mu$ as long as $B^\dagger\phi_\mu\neq0$. For a non-zero $\lambda_\mu$, $B^\dagger\phi_\mu$ cannot be zero (it leads to $\lambda_\mu=0$). Using these observations and that $H,H^\dagger$ have the same spectrum, we then arrive at the conclusion that the spectra of $H$ and $H_e$ are the same, but one cannot eliminate the possibility that $H$ has EP(s).  

\end{itemize}

As we mentioned in the introduction, our construction of a non-Hermitian $H$ with a real spectrum can also be found in a system of coupled frictionless harmonic oscillators with different masses $m_i$'s:
\be
m_i\ddot{x}_i = -2kx_i + k(x_{i-1} + x_{i+1}).
\ee
Here ${x}_i$ is the displacement of the $i$th mass, and all masses are connected sequentially by springs with the same spring constant $k\in\mathbb{R}$. If we define the dynamical matrix $M$ as 
\be
\ddot{\bm{x}} = M\bm{x}\label{eq:M}
\ee
where $\bm{x} = [x_1,x_2,\ldots,x_N]^T$, then $M$ has the same form $AM_0$ as in our construction, where $A$ is a diagonal and positive-definite matrix with elements $a_j = m_j^{-1}$ and $M_0$ is Hermitian and tri-diagonal, with $-2k$ and $k$ on its main diagonal and upper/lower diagonals. Clearly, the system is dissipationless and Hermitian, and hence the harmonic solutions  $\bm{x}(t)=\bm{x}(0) e^{-i\omega_\mu t}$ have real-valued eigenfrequencies $\omega_\mu$'s. Consequently, the eigenvalues of $M$, given by $-\omega_\mu^2$, are all real, despite that $M$ is non-Hermitian if $m_i$'s are different (e.g., $M_{1,2} = k/m_1\neq M_{2,1}^* = k/m_2$). 

% We also note that to define an effective ``Hamiltonian'' for Eq.~(\ref{eq:M}), one typically rewrites the latter in the form with only first-order time derivatives:
% \be
% i\frac{d}{dt}
% \begin{pmatrix}
% \dot{\bm{x}} \\
% \bm{x}
% \end{pmatrix}
% = 
% i\begin{pmatrix}
% 0 & M \\
% \bm{1} & 0
% \end{pmatrix}
% \begin{pmatrix}
% \dot{\bm{x}} \\
% \bm{x}
% \end{pmatrix}
% \equiv
% iH 
% \begin{pmatrix}
% \dot{\bm{x}} \\
% \bm{x}
% \end{pmatrix}.
% \label{eq:M2}
% \ee
% Due to its zero diagonal blocks, this effective Hamiltonian $H$ has non-zero chiral symmetry \cite{chiral}, i.e.,
% \be
% \{H,C\}=HC-CH=0, 
% \ee
% where
% \be
% C = \begin{pmatrix}
% \bm{1} &\\
% & -\bm{1}
% \end{pmatrix}\label{eq:C}.
% \ee
% As a result, all eigenvalues of $H$ appear in pairs that satisfy $\omega_{\mu}=-\omega_{\nu}$. The corresponding eigenstates are related by $U_{\mu} = CU_\nu$, and by denoting $U_{\mu}=[\phi_{\mu}, \psi_{\mu}]^T$, Eq.~(\ref{eq:M2}) leads to
% \be
% M\psi_{\mu} = \omega_{\mu}\phi_{\mu},\quad\phi_{\mu} = \omega_{\mu} \psi_{\mu},
% \ee
% or simply,
% \be
% M\psi_{\mu} = \omega_{\mu}^2\psi_{\mu}.
% \ee
% In other words, the eigenfrequencies $\omega_{\mu}$'s are just the square roots of the eigenvalues $\omega_\mu$'s of $M$. 

In this mechanical system, although $M$ has a non-Hermitian form, the system itself is still Hermitian, and one can directly measure the real spectrum of the oscillation modes. This procedure, however, is not generally applicable in a non-Hermitian system. 
In order to probe a non-Hermitian system with a real spectrum, one can then shift the energy of this system such that all energy levels acquire the same imaginary part or linewidth. This linewidth can then be measured either in a passive or active system, which has been done, for example, in PT-symmetric dimers \cite{NPReview,RMP,Longhi}. We expect the same experimental procedure to resolve the real spectrum in our proposed non-Hermitian Hamiltonian, where the reduced lasing threshold of the selective non-Hermitian skin mode can also be verified by comparing to a system on the same on-chip semiconductor platform with the standard non-Hermitian skin effect \cite{Gao}. Such lasers operating in the single or few modes regimes tend to be stable dynamically, thanks to the saturable nonlinearity of the gain medium.

We also stress that even when a non-Hermitian system has a real spectrum, it is our opinion that it should still be considered as a subsystem that exchanges particles and energy with its environment. Therefore, one does not expect the conservation of probability to hold. 

Nevertheless, there are other forms of generalized conservation laws that can be derived in non-Hermitian systems. One simple example is the conservation of probability or optical flux in a one-dimensional system. In a Hermitian system, this conservation law states that $R+T=1$, i.e., the summation of the reflected flux and the transmitted one is conserved. In a non-Hermitian PT-symmetric system, this conservation law evolves to $\sqrt{R_LR_R}=|1-T|$ \cite{conservation}, where $T$ is the still reciprocal transmittance from either the left or right side of the system which can now be larger than 1, while the reflectance from the left $R_L$ and that from the right $R_R$ can be different. In the Hermitian limit, $T$ is always smaller than 1 and $R_R=R_L\equiv R$, and we recover $R+T=1$. Another example is the generalized Ehrenfest theorem defined with respect to a non-Hermitian inner product $(A)\equiv\psi^TA\psi$. When the linear operator $A$ is not explicitly time-dependent, i.e. $(dA/dt)=0$, then it can be shown that $(A)$ is a constant of motion when $A$ defines a pseudo-chirality of the system, i.e., $H^T = -AHA^{-1}$ \cite{pseudoChirality}.

\section{Conclusion}

In summary, we have presented a systematic approach to constructing non-Hermitian Hamiltonians with a real spectrum beyond a simple similar transformation, which can have embedded EPs at $\omega_\mu=0$ depending on whether $A$ has zero eigenvalue(s). A simple spectral shift can also realize an EP at any desired energy or frequency. Our approach utilizes the product of a Hermitian Hamiltonian $H_0$ and a positive semi-definite matrix $A$. %As such, the resulting non-Hermitian Hamiltonian $H$ has a different spectrum than the underlying Hermitian Hamiltonian $H_0$. 
Despite bearing similarity to the polar decomposition of an arbitrary square matrix, i.e., as a product of a unitary matrix $U$ and a positive semi-definite matrix $A'$, here $A$ and $A'$ of $H$ are different: $A'$ is given by $(H^\dagger H)^{1/2}=(A H_0^2A)^{1/2}$ instead.

The positive semi-definite matrix $A$ does not need to be diagonal, and when it is, we show that our approach can lead to a selective non-Hermitian skin effect, where the zero mode of the system is a skin mode while the non-zero modes still reside in the bulk. 
We have further shown that the formation of this selective non-Hermitian skin mode is due to two complement mechanisms, i.e., an imaginary gauge transformation and a non-uniform effective coupling strength, each contributing the same exponential factor to the localized spatial profile. Due to the reduced loss to the photonic environment at the non-reciprocal coupling junctions, the lasing threshold of this selective non-Hermitian skin mode is also much lower than its counterpart with the same spatial profile in the standard non-Hermitian skin effect, when we pump at the boundary where they are localized. Our approach may also find applications in constructing and understanding other non-Hermitian matrix operators, including but not limited to the coupled frictionless harmonic oscillators we have discussed. 

\section{Methods}

\noindent \textbf{Perturbation theory.} Below we explain the different amplitudes of the zero modes at their lasing thresholds in $H$ and $H''$, shown in Fig.~\ref{fig:laser}(a) of the main text. We treat the localized pump $\gamma_1$ in the leftmost cavity as the small perturbation parameter, which is justified given that $\gamma_1$ is on the order of the loss $\kappa_0$, which is much smaller than the couplings in this case.

We denote the effective Hamiltonian with the pump by
\be
H_a=H_p+i\gamma_1H_\gamma,
\ee
where $H_p$ is the passive Hamiltonian representing $H$ or $H''$ with the cavity loss $\kappa_0$. The perturbation $H_\gamma$ has a single non-zero element 1 at the upper left corner (i.e., the diagonal element of the leftmost cavity). The right and left eigenstates of $H_p$ are denoted by $|\psi_\mu\rangle$ and $\langle\tilde{\psi}_\nu|$, and they satisfy the biorthogonal relation
\be
\langle\tilde{\psi}_\nu|\psi_\nu\rangle = \delta_{\nu\mu} \label{eq:biortho}
\ee
because the system is away from an EP of $H_p$. 
 
Similar to the perturbation theory in (Hermitian) quantum mechanics, the first-order correction to the energy is given by
\be
\omega_\mu^{(1)}=i\gamma_1\langle\tilde{\psi}_\mu|H_\gamma|\psi_\mu\rangle\equiv i\gamma_1 H_{\gamma,\mu\mu}.
\ee
For the passive zero mode with $\gamma_1=0$, its wave function is real (zero) when $j$ is odd (even), which holds for both the corresponding left and right eigenstates. Therefore, the first-order energy correction $\omega_0^{(1)}$ to the zero mode is imaginary, meaning that the zero mode does not change its (real) frequency as we pump the system. This is the non-Hermitian zero mode \cite{NHFlatband_PRL,NHFlatband_PRJ} warranted by the NHPH symmetry we have mentioned in the main text.

\begin{figure}[t]
\includegraphics[clip,width=\linewidth]{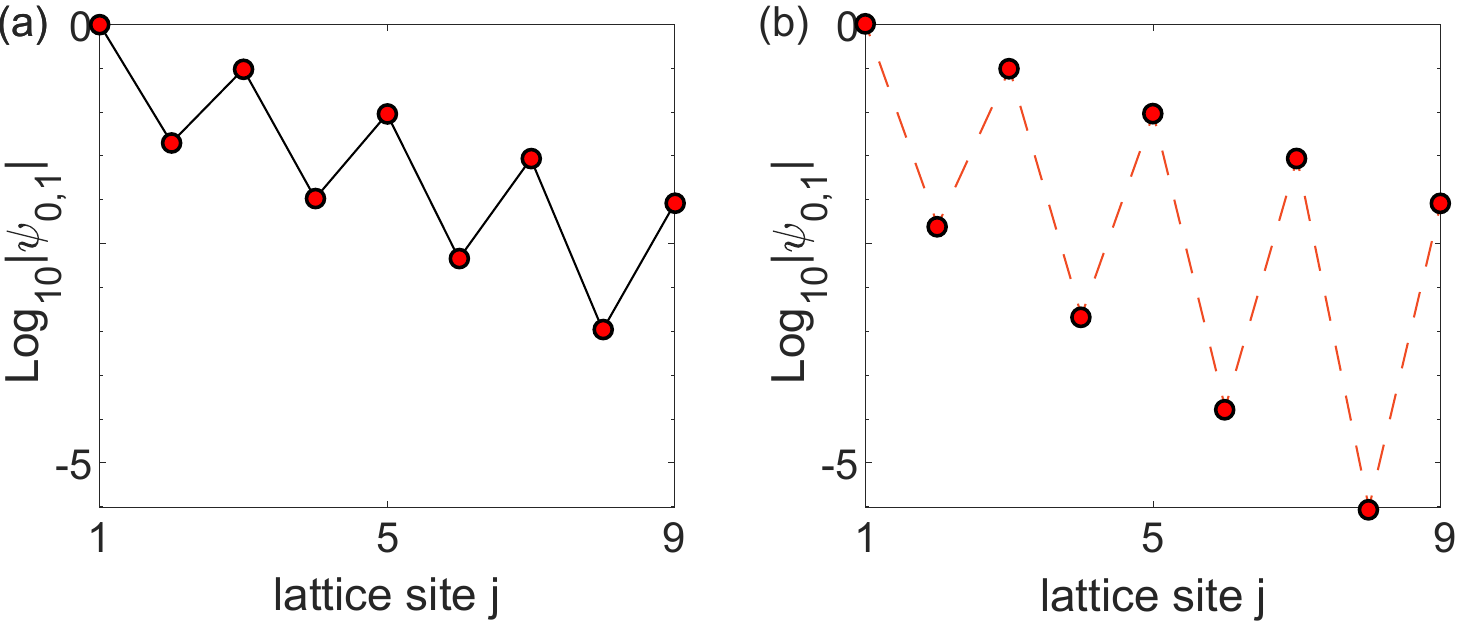}
\caption{\textbf{Non-Hermitian perturbation results} (dots) for the zero mode at threshold in the selective (a) and standard (b) non-Hermitian skin effects. Lines in (a) and (b) are the same as in Fig.~\ref{fig:laser}(a) of the main text. } \label{fig:pert}
\end{figure}

In the meanwhile, the first-order correction to the wave function is given by
\be
|\psi_\mu^{(1)}\rangle = i\gamma_1\sum_{\nu\neq\mu}\frac{H_{\gamma,\nu\mu}}{\omega_\mu-\omega_\nu}|\psi_\nu\rangle.
\ee
The NHPH symmetry dictates 
\be
\psi_{\nu,j} = (-1)^{j-1}\psi_{\nu',j},\quad \tilde{\psi}_{\nu,j} = (-1)^{j-1}\tilde{\psi}_{\nu',j},\quad \omega_\nu = -\omega_{\nu'}^* \nonumber
\ee
for each pair of non-zero modes, which lead to  
\be
H_{\gamma,\nu 0} = H_{\gamma,\nu' 0},\quad \omega_0-\omega_\nu = -(\omega_0-\omega_{\nu'})\in\mathbb{R}, 
\ee
and in turn
\be
\psi_{0,j}^{(1)} = 
\begin{cases}
0, & \quad(j=\text{odd}) \\
-2i\gamma_1 \sum_{\nu=0}^{(N-1)/2} \frac{H_{\gamma,\nu 0}}{\re{\omega_\nu}}\psi_{\nu,j}.& \quad(j=\text{even}) \\
\end{cases}\nonumber	
\ee
The first line of this expression then explains the (almost) identical wave functions of the zero modes at threshold in odd-numbered cavities shown in Fig.~\ref{fig:laser}(a), which are essentially given by the underlying passive zero modes in $H$ and $H''$ that are identical. The second line captures the significant difference between the zero modes in even-numbered cavities in $H$ and $H''$ [see Fig.~\ref{fig:pert}], which leads to the one-order of magnitude difference in their loss to the photonic environment at the coupling junctions [Fig.~\ref{fig:laser}(b) in the main text], and in turn, their contrasting lasing thresholds.

\acknowledgements

This work is supported by the National Science Foundation (NSF) under Grants No. PHY-1847240 and No. DMR-2326698.


\begin{thebibliography}{99}

\bibitem{Gamow} G. Gamow, Zur quantentheorie des atomkernes,
Z. Phys. \textbf{51}, 204 (1928).

\bibitem{SALT_Science} H. E. T\"ureci, L. Ge, S. Rotter, and A. D. Stone, Storng interactions in multimode random lasers,
Science {\bf 320}, 643 (2008).

\bibitem{SALT_PRA} L. Ge, Y. D. Chong, and A. D. Stone, Steady-State Ab Initio Laser Theory: Generalizations and Analytic Results, 
Phys. Rev. A \textbf{82}, 063824 (2010).

\bibitem{Bender} C. M. Bender and S. Boettcher, Real Spectra in Non-Hermitian Hamiltonians Having PT Symmetry, 
Phys. Rev. Lett. \textbf{80}, 5243 (1998).


\bibitem{Mostafazadeh2} A. Mostafazadeh, Pseudo-Hermiticity versus PT-Symmetry. II. A Complete Characterization of Non-Hermitian Hamiltonians with a Real Spectrum, 
J. Math. Phys. \textbf{43}, 2814 (2002).

\bibitem{NPReview} L. Feng, R. El-Ganainy, and L. Ge, Non-Hermitian photonics based on parity-time symmetry,
Nat. Photon. \textbf{11}, 752--762 (2017).
\bibitem{RMP} V. V. Konotop, J. Yang, and D. A. Zezyulin, Nonlinear waves in PT-symmetric systems,
Rev. Mod. Phys. \textbf{88}, 035002 (2016).
\bibitem{Longhi} S. Longhi, Parity-Time Symmetry Meets Photonics: A New Twist in Non-Hermitian Optics, 
EPL \textbf{120}, 64001 (2017).

\bibitem{EPsensing1} W. Chen, S. K. Ozdemir, G. Zhao, J. Wiersig, L. Yang, Exceptional points enhance sensing in an optical microcavity, Nature \textbf{548}, 192 (2017).
\bibitem{EPsensing2} Y.-H. Lai, Y.-K. Lu, M.-G. Suh, Z. Yuan, and K. Vahala, Observation of the exceptional-point-enhanced Sagnac effect, Nature \textbf{576}, 65 (2019).
\bibitem{EPsensing3} H. Hodaei et al., Enhanced sensitivity at higher-order exceptional points, Nature \textbf{548}, 187 (2017).

\bibitem{DiracEP_PRL} J. H. D. Rivero, L. Feng, and L. Ge, Imaginary Gauge Transformation in Momentum Space and Dirac Exceptional Point, 
Phys. Rev. Lett. \textbf{129}, 243901 (2022).

\bibitem{DiracEP_PRB} J. H. D. Rivero, L. Feng, and L. Ge, Analysis of Dirac Exceptional Points and Their Isospectral Hermitian Counterparts, 
Phys. Rev. B \textbf{107}, 104106 (2023).


\bibitem{Martinez} B. M. Villegas-Martínez, F. Soto-Eguibar, S. A. Hojman, F. A. Asenjo, and H. M. Moya-Cessa, Non-Unitary Transformation Approach to $\mathcal{PT}$ Dynamics, 
arXiv:2201.06536.

\bibitem{Hatano} N. Hatano and D. R. Nelson, Localization Transitions in Non-Hermitian Quantum Mechanics,
Phys. Rev. Lett. \textbf{77}, 570--573 (1996).

\bibitem{Hatano2} N. Hatano and D. R. Nelson, Vortex Pinning and Non-Hermitian Quantum Mechanics, 
Phys. Rev. B \textbf{56}, 8651 (1997).

\bibitem{Gao} Z. Gao, X. Qiao, M. Pan, S. Wu, J. Yim, K. Chen, B. Midya, L. Ge, and L. Feng, Two-Dimensional Reconfigurable Non-Hermitian Gauged Laser Array, 
Phys. Rev. Lett. \textbf{130}, 263801 (2023).
\bibitem{acoustic} X. Zhang, Y. Tian, J.-H. Jiang, M.-H. Lu, and Y.-F. Chen, Observation of higher-order non-Hermitian skin effect, 
Nat. Commun. \textbf{12}, 5377 (2021).
\bibitem{mechanical} W. Wang, X. Wang, and G. Ma, Non-Hermitian morphing of topological modes, 
Nature \textbf{608}, 50 (2022).


\bibitem{Bronson} R. Bronson and G. B. Costa, \textit{Matrix methods}, 4th ed., (Academic Press, San Diego, CA, 2020).

\bibitem{chiral} J. D. H. Rivero and L. Ge, Chiral Symmetry in Non-Hermitian Systems: Product Rule and Clifford Algebra, 
Phys. Rev. B \textbf{103}, 014111 (2021).

\bibitem{harmonic} P. Leboeuf and S. Moulieras, Superfluid Motion of Light, 
Phys. Rev. Lett. \textbf{105}, 163904 (2010).


\bibitem{SSH} W. P. Su, J. R. Schrieffer, and A. J. Heeger, Solitons in polyacetylene, 
Phys. Rev. Lett. \textbf{42}, 1698 (1979).
\bibitem{Kagome1} M. Ezawa, Higher-Order Topological Insulators and Semimetals on the Breathing Kagome and Pyrochlore Lattices, 
Phys. Rev. Lett. \textbf{120}, 026801 (2018).

\bibitem{microcavity1} \textit{Optical Processes in Microcavities}, edited by R. K. Chang and A. J. Campillo, Advanced Series in Applied Physics (World Scientific, Singapore, 1996). 
\bibitem{microcavity2} \textit{Optical Microcavities}, edited by K. J. Vahala, Advanced Series in Applied Physics (World Scientific, Singapore, 2004). 
\bibitem{microcavity3} H. Cao and J. Wiersig, Dielectric microcavities: Model systems for wave chaos and non-Hermitian physics, Rev. Mod. Phys. \textbf{87}, 61 (2015).

\bibitem{anomalousTH} L. Ge, Z. Gao, and L. Feng, Non-Hermitian Gauged Laser Arrays with Localized Excitations: Anomalous Threshold and Generalized Principle of Selective Pumping, 
Phys. Rev. B \textbf{108}, 104111 (2023).




\bibitem{Ge_PRA_2017b} L. Ge, K. G. Makris, and L. Zhang, Optical Fluxes in Coupled PT-Symmetric Photonic Structures, 
Phys. Rev. A \textbf{96}, 023820 (2017).

\bibitem{zeromodeLaser} L. Ge, Symmetry-protected zero-mode laser with a tunable spatial profile,
Phys. Rev. A \textbf{95}, 023812 (2017).

\bibitem{NHFlatband_PRL} B. Qi, L. Zhang and L. Ge, Defect states emerging from a non-Hermitian flat band of photonic zero modes,
Phy. Rev. Lett. \textbf{120}, 093901 (2018).
\bibitem{NHFlatband_PRJ} L. Ge, Non-Hermitian lattices with a flat band and polynomial power increase,
Photon. Res. \textbf{6}, A10--A17 (2018).

\bibitem{linearLoc} B. Qi and L. Ge, Linear Localization of Zero Modes in Weakly Coupled Non-Hermitian Reservoirs, Adv. Phys. Research \textbf{2}, 2300066 (2023).

\bibitem{Jordan} H.-Z. Chen et al., Revealing the Missing Dimension at an Exceptional Point, 
Nat. Phys. \textbf{16}, 571 (2020).

\bibitem{conservation} L. Ge, Y. D. Chong, and A. D. Stone, Conservation Relations and Anisotropic Transmission Resonances in One-Dimensional PT-Symmetric Photonic Heterostructures, 
Phys. Rev. A \textbf{85}, 023802 (2012).
\bibitem{pseudoChirality} J. D. H. Rivero and L. Ge, Pseudochirality: A Manifestation of Noether’s Theorem in Non-Hermitian Systems, 
Phys. Rev. Lett. \textbf{125}, 083902 (2020).



\end{thebibliography}
\end{document}